\documentclass{article}

\usepackage[aux]{rerunfilecheck}
\usepackage{amsmath}
\usepackage{amssymb}
\usepackage{mathtools}
\usepackage[numbers,sort&compress]{natbib}
\usepackage[autostyle]{csquotes}
\usepackage{xfrac}
\usepackage{float}
\usepackage{graphicx}
\usepackage{color}
\usepackage{bbm} 
\usepackage{placeins} 
\usepackage{booktabs}

\newcommand{\rvec}[2]{\ensuremath{\begin{pmatrix}#1 & #2\end{pmatrix}}}

\title{\bf Searching New Particles at Neutrino Telescopes with Quantum-Gravitational Decoherence}

\author{Dominik Hellmann$^1$, Heinrich P\"as$^1$, Erika Rani$^{1,2}$
\smallskip
\\
{\it $^1$ Fakult\"at f\"ur Physik,
Technische Universit\"at Dortmund,
Germany}
\\
{\it $^2$ UIN Maulana Malik Ibrahim Malang, Indonesia}
}

\begin{document}

\maketitle

\begin{abstract}
We discuss the interplay of wave packet decoherence and decoherence induced by quantum gravity via interactions with spacetime foam for high energy astrophysical neutrinos.
In this context we point out a compelling consequence of the expectation that quantum gravity should break global symmetries, namely
that quantum-gravity induced decoherence may not only
be the most sensitive probe for quantum properties of spacetime,
but also can provide both a powerful tool for the search for new particles, including totally decoupled backgrounds interacting only gravitationally, and at the same time a window into the intricacies of black hole information processing.
\end{abstract}

\section{Introduction}
The search for quantum effects of gravity and the exploration of the quantum-to-classical transition belong to the most exciting frontier areas in fundamental physics. Yet the interplay between both kinds of effects in quantum-gravitational decoherence is addressed in the literature only rarely. In this letter we point out that quantum-gravitational decoherence may provide groundbreaking opportunities for a third important frontier area of fundamental physics, namely the search for new, electrically neutral particles, including fermionic dark matter, new ``sterile'' neutrino degrees of freedom beyond the Standard Model of Particle Physics and totally decoupled backgrounds interacting only gravitationally.

Decoherence describes the loss of coherence of a quantum system due to interactions with an environment and
is generally accepted as one if not the defining aspect of the quantum-to-classical transition \cite{Zeh:1970zz}.
The concept that quantum-gravity could induce decoherence of elsewise isolated quantum-systems is based on John Wheeler's idea that quantum spacetime should exhibit a foamy structure comprised of virtual black holes on small scales of order $M_{\rm P}^{-1}$, with $M_{\rm P}$ being the Planck mass (\cite{Wheeler:1955zz}, for a recent review see \cite{AmelinoCamelia:2008qg}). The dynamical properties of spacetime foam may act then as a decoherence-inducing environment.
Such decoherence triggered by interactions with spacetime foam has an interesting property, namely that it is expected to violate global symmetries. The idea goes back to classical papers on black-hole evaporation by Hawking (e.g. \cite{Hawking:1974sw}) and Page \cite{Page:1980qm}: According to General Relativity, a black hole is fully characterized by mass, angular momentum and charge, all other information such as global quantum numbers (flavor, baryon number, lepton number) is supposedly lost (also known as “No Hair Theorem”). Thus the interaction with a black hole should violate the unitarity of quantum mechanics.
If gravity is quantized, this may be observable in interactions with the virtual black hole background present in spacetime foam, see e.g. \cite{Anchordoqui:2006xv} or \cite{Witten:2017hdv} for a review. These heuristic conjectures have recently been confirmed with holographic arguments in the context of the AdS/CFT correspondence
\cite{Harlow:2018jwu}.
Ellis, Hagelin, Nanopoulos and Srednicki described such effects with a sink term in the Liouville-von-Neumann equation giving rise to gravitational-induced decoherence in vacuo (\cite{Ellis:1983jz}, see also
\cite{Ellis:1992dz,Huet:1994kr}). An application to neutrino oscillations has been worked out e.g. in
\cite{Liu:1997zd,Chang:1998ea,Benatti:2000ph,Gago:2002na,Barenboim:2006xt}.

As a consequence of the breaking of global symmetries by quantum gravity, this type of decoherence is typically considered to be independent on flavor mixing. For example, a beam of astrophysical neutrinos will be distributed democratically over all flavors, implying that flavor ratios probed in neutrino telescopes such as e.g. IceCube may be sensitive to such effects. As has been pointed out more than two decades ago by one of us and collaborators \cite{KlapdorKleingrothaus:2000fr} and later developed e.g. in \cite{Hooper:2004xr,Hooper:2005jp,Anchordoqui:2005gj,Stuttard:2020qfv}, the study of astrophysical neutrinos provides an extremely sensitive probe on such kind of effects.

There are at least three good reasons to revisit this interesting phenomenon now:

\begin{enumerate}

\item
The mounting cosmological evidence for dark matter in the universe combined with the fact that no new particles have been found at the Large Hadron Collider (LHC) so far.

\item
The recent developments in the research of quantum gravity that brought the question how black holes process information to the centerstage of attention \cite{tHooft:1984kcu,Susskind:1993if,Almheiri:2012rt}
and inspired a wealth of interesting works about the relations between spacetime and quantum information
that lack however concrete possibilities to be probed experimentally.

\item
The discovery of PeV scale extragalactic neutrinos in the IceCube neutrino telescope
\cite{Aartsen:2013jdh,IceCube:2018cha,Abbasi:2021bvk}.

\end{enumerate}

In this paper, we thus study the transition from standard wave packet decoherence to the hypothetical quantum-gravitational decoherence. In this context we focus on the energy dependence of the mechanism and point out a curious phenomenon that we believe has been overlooked so far,
namely that the state space must be extended if there exist neutral fermions beyond the three SM neutrinos. By deriving the
consequences of this extension we show that
a global symmetry breaking decoherence process provides a sensitive portal to fermionic hidden sectors 
of the universe that feature no unbroken gauge quantum numbers
and thus a powerful tool to search for neutral fermions, including undiscovered neutrino flavors, dark matter particles and totally decoupled sectors that interact only gravitationally with the Standard Model particles.

\section{Decoherence in the 2$\nu$ Framework}
We start with a Lindblad equation describing the density matrix \(\varrho(t)\) of an open quantum system being exposed to both wave packet separation and quantum-gravitational decoherence,
\begin{align}
    \label{eq:Lindblad}
    \frac{\mathrm{d}}{\mathrm{d} t}\varrho(t) = -i[H,\varrho(t)] - \frac{1}{L_{\mathrm{coh}}} \left(1 - \hat{D}\right)\varrho(t) - \mathcal{G} \varrho(t).
\end{align}
Here, the first term describes the standard von-Neumann time evolution of an undisturbed quantum
system (giving rise to the typical flavor oscillation for \(L\ll L_{\rm coh}\))
with the Hamiltonian given in the mass basis as
\begin{align}
    H &= \begin{pmatrix}
        0 & 0 \\
        0 & \frac{\Delta m^2}{2E}
    \end{pmatrix},
\end{align}
and the following terms describe the loss of unitarity due to wave packet~\cite{Akhmedov:2014ssa}
and quantum-gravitational decoherence, respectively.
The term corresponding to the effect of wave packet separation is chosen such that it gives rise to a simple exponential damping of the coherence between neutrino mass eigenstates, assuming Lorentzian wave packets~\cite{Akhmedov:2009rb}.\footnote{Since the coherence damping is always of exponential type~\cite{Akhmedov:2009rb}, the concrete wave packet shape does not affect the conclusions of this work.} 
To illustrate the basic features of the phenomenon under investigation, we consider here a 2-neutrino scenario
where \(\Delta m^2\) is the mass squared difference of the two states involved.
In the following, we use the abreviation \(\Delta := \sfrac{\Delta m^2}{2E}\) for a less cluttered notation.
Furthermore, \(\hat{D}\) is an operator projecting out the diagonal elements of a \(2 \times 2\) matrix and
\begin{align}
    L_{\mathrm{coh}} = \sigma_x \frac{2E^2}{\Delta m^2}
\end{align}
is the coherence length~\cite{Kersten:2015kio} of the neutrino system of energy \(E\) and initial wave packet size \(\sigma_x\).\\

The action of the operator \(\mathcal{G}\) is defined as~\cite{Liu:1997zd}
\begin{align}
    \label{eq:QG_Dissipator}
    \mathcal{G} \varrho &= 2 h_{ij}\sigma_i\varrho_j,\\
    h &=    \begin{pmatrix}
                0 & 0 & 0 \\
                0 & \alpha & \beta \\
                0 & \beta & \gamma \\
            \end{pmatrix},
\end{align}
where \(\alpha, \beta, \gamma \propto \sfrac{E^{n}}{M_{\rm P}^{n-1}}\) are the quantum-gravity decoherence parameters and \(n\) determines their energy dependence.
From now on we use a single parameter $\gamma = \alpha$, i.e. the universal interaction rate of the system with the spacetime foam, to parametrize the exponential damping of the density matrix components which is mainly controlled by the diagonal entries of $h_{ij}$.
Furthermore, we make the common choice of $n = 2$ motivated by several Planck scale models~\cite{Ellis:1996bz,Ellis:1997jw,BENATTI199958} enabling us to employ the upper bounds derived in~\cite{Lisi:2000zt} and used in~\cite{Stuttard:2020qfv}.\\
By writing Eq.~\eqref{eq:QG_Dissipator}, we already expanded the result in the basis of the Pauli matrices.
If we do this for all terms in~\eqref{eq:Lindblad}, we arrive at a differential equation for the components \(\varrho_j\) of the density matrix
\begin{align}
    \frac{\mathrm{d}}{\mathrm{d} t}\vec{\varrho} &= \mathcal{L}\vec{\varrho} \label{eq:DGL},
\end{align}
with
\begin{align}
    \mathcal{L} &= \begin{pmatrix}
        0 & 0               & 0                         & 0 \\
        0 & -\frac{1}{L_{\mathrm{coh}}}  & \Delta                    & 0 \\
        0 & -\Delta         & -2\gamma - \frac{1}{L_{\mathrm{coh}}}  & -2\beta \\
        0 & 0               & -2\beta                   & -2\gamma \\
    \end{pmatrix}\,.
\end{align}
Eq.~\ref{eq:DGL}
can be easily solved using the assumption \(\vert\beta\vert \ll \vert\gamma\vert\),
implying that \(\mathcal{L} = S J S^{-1}\)
is diagonalized by the matrix
\begin{align}
    S &=    \begin{pmatrix}
                1 & 0 & 0 & 0\\
                0 & 0 & \frac{-\gamma+i\sqrt{\Delta^2-\gamma^2}}{\Delta} & \frac{-\gamma-i\sqrt{\Delta^2-\gamma^2}}{\Delta}\\
                0 & 0 & 1 & 1\\
                0 & 1 & 0 & 0\\
            \end{pmatrix}
\end{align}
and its inverse, leading to
\begin{align}
    J &= \mathrm{diag}\left(0,-2\gamma, -\gamma - \frac{1}{L_{\mathrm{coh}}}-i\sqrt{\Delta^2-\gamma^2}, -\gamma - \frac{1}{L_{\mathrm{coh}}}+i\sqrt{\Delta^2-\gamma^2}\right)\,.
\end{align}
Using this decomposition, we can give an explicit solution to Eq.~\eqref{eq:DGL}, i.e.
\begin{align}
    \vec{\varrho}(t) &= S\exp\left(Jt\right)S^{-1}\vec{\varrho}(0)\,.
\end{align}
Assuming the initial neutrino to be an electron neutrino,
\begin{align}
    \vec{\varrho}(0) = \left(\frac{1}{2},\frac{1}{2}\sin(2\theta),0,\frac{1}{2}\cos(2\theta)\right)^T\,,
\end{align}
we obtain the evolution of the density matrix as
\begin{align}
    \vec{\varrho}(t) &= \begin{pmatrix}
        \frac{1}{2} \\
        \frac{1}{2}e^{-\left(\gamma+\frac{1}{L_{\mathrm{coh}}}\right)t}\sin(2\theta)\left\{\cos\left(\omega t\right) + \frac{\gamma}{\omega}\sin\left(\omega t\right)\right\}\\
        -\frac{1}{2}e^{-\left(\gamma+\frac{1}{L_{\mathrm{coh}}}\right)t}\sin(2\theta)\frac{\Delta}{\omega}\sin\left(\omega t\right) \\
        \frac{1}{2}\cos(2\theta)e^{-2\gamma t}
    \end{pmatrix},
\end{align}
where
    \(\omega := \sqrt{\Delta^2-\gamma^2}\) and
\(\theta\) is the neutrino mixing angle.
From this, we can calculate the probability for detecting an electron neutrino (\(P_{ee}(L)\)) or a differently flavored neutrino (\(P_{e(\mu\tau)}(L)\)) at a certain baseline \(L \simeq t\)
\begin{align}
    P_{ee}(L) &= \frac{1}{2} + \frac{1}{2}\cos^2(2\theta)e^{-2\gamma L}\nonumber\\
                &+\frac{1}{2}\sin^2(2\theta)e^{-\left(\gamma+\frac{1}{L_{\mathrm{coh}}}\right)L}\left\{\cos\left(\omega L\right)+ \frac{\gamma}{\omega}\sin\left(\omega L\right)\right\}\,,\label{eq:result1}\\
    P_{e(\mu,\tau)}(L)&= 1 - P_{ee}(L)\label{eq:result2}\,.
\end{align}
The oscillation formulae for quantum-gravitational decoherence in the three neutrino case have been
presented in~\cite{Gago:2002na,Barenboim:2006xt}.
In the limiting case of \(L_{\mathrm{coh}} \ll L \ll \frac{1}{\vert\gamma\vert}\), the
exponential factor \(e^{-2\gamma L}\) is still close to unity
and the usual results of wave packet decoherence are recovered:
\begin{align}
    P_{ee}(L) &= 1 - \frac{1}{2}\sin^2(2\theta)\,,\\
    P_{e(\mu\tau)}(L) &= \frac{1}{2}\sin^2(2\theta)\,.
\end{align}
If \( L\) approaches
\(\frac{1}{\vert\gamma\vert}\), i.e. for large baselength and high energies, we obtain
\begin{align}
    P_{ee}(L) &= \frac{1}{2} + \frac{1}{2}\cos^2(2\theta)e^{-2\gamma L}\,,\\
    P_{e(\mu\tau)}(L) &= \frac{1}{2} - \frac{1}{2}\cos^2(2\theta)e^{-2\gamma L}\,.
\end{align}

For illustration, we consider electron neutrinos originating from neutron decay oscillating into a maximally mixed superposition of \(\nu_\mu\) and \(\nu_\tau\),
\(\Delta m^2 = \Delta m_{\mathrm{sol}}^2 = 7.53\cdot 10^{-5}\;\mathrm{eV^2}\) and \(\sin^2(2\theta) = \sin^2(2\theta_{\mathrm{sol}}) \approx 0.85\)~\cite{Zyla:2020zbs}. For the quantum gravity parameter we adopt the common~\cite{Stuttard:2020qfv,Lisi:2000zt} parametrization \(\gamma = \xi \sfrac{E^2}{M_{\rm P}}\) with \(\xi = 10^{-28}\), many orders of magnitudes below the upper bounds obtained in Refs~\cite{Stuttard:2020qfv,Lisi:2000zt} and the more natural expectation of \(\xi \lesssim \mathcal{O}(1)\) if the underlying energy scale is the Planck scale.
This parameter choice demonstrates the ground-breaking sensitivity of astrophysical neutrinos, even for extremely large coherence lengths \(L_{\mathrm{coh}}^{\mathrm{QG}}\propto \sfrac{1}{\gamma}\).
Finally, we adopt \(\sigma_x = 10^{-11}\;\mathrm{cm}\)~\cite{Kersten:2015kio} for the initial neutrino wave packet size, corresponding to neutrinos produced in very short lived processes.

\begin{figure}
    \centering
    \includegraphics[width=\textwidth]{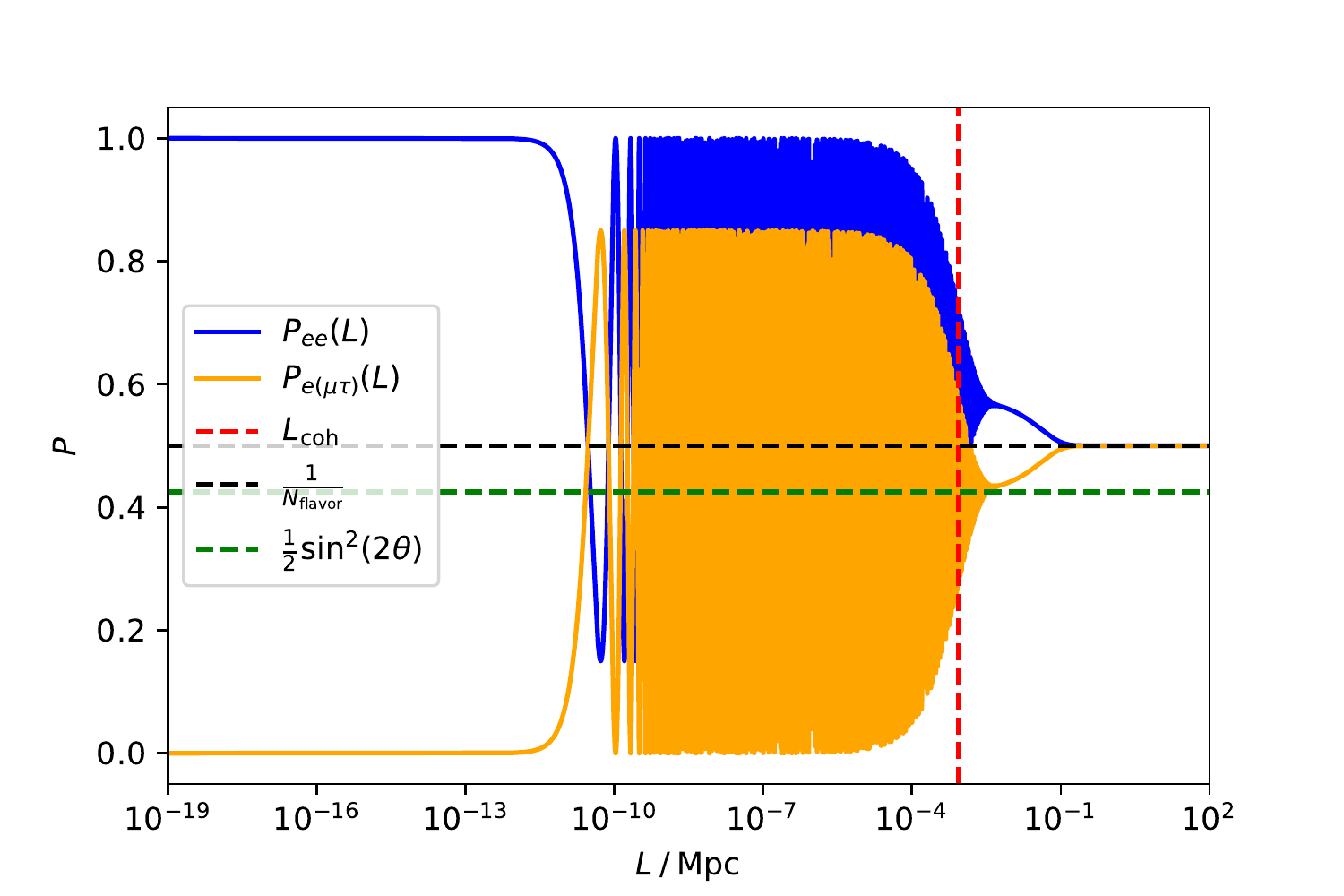}
    \caption{Probabilities for an initial electron neutrino of \(E = 100\;\mathrm{TeV}\) to oscillate into an electron neutrino and a differently flavored neutrino, which is either the tau or the muon neutrino, respectively. The blue curve depicts the probability \(P_{ee}(L)\), while the orange curve represents the probability \(P_{e(\mu\tau)}(L)\). Furthermore, we show dashed, horizontal lines for the asymptotic values of \(\sfrac{1}{2}\sin^2(2\theta)\) (green) and \(\sfrac{1}{N_{\mathrm{flavor}}}\) (black) and a dashed vertical line for the position of the coherence length \(L_{\mathrm{coh}}\)(red).}
    \label{fig:result_plot_L}
\end{figure}

Fig.~\ref{fig:result_plot_L}  highlights the sensitivity of astrophysical neutrinos to quantum-gravi\-ta\-tional decoherence by plotting
the baseline dependence of the oscillation probabilities from equations~\eqref{eq:result1} and~\eqref{eq:result2} for neutrino energies of  \(E = 100\; \mathrm{TeV}\). As can be seen, the no-oscillation regime at small baselines
is followed by standard neutrino oscillations at medium baselines that are supplanted first by standard wave packet decoherence and finally by hypothetical quantum gravity decoherence, implying a limiting oscillation probability of
$1/n$ (=0.5 in the present example) for all flavors involved.

\begin{figure}
    \centering
    \includegraphics[width=\textwidth]{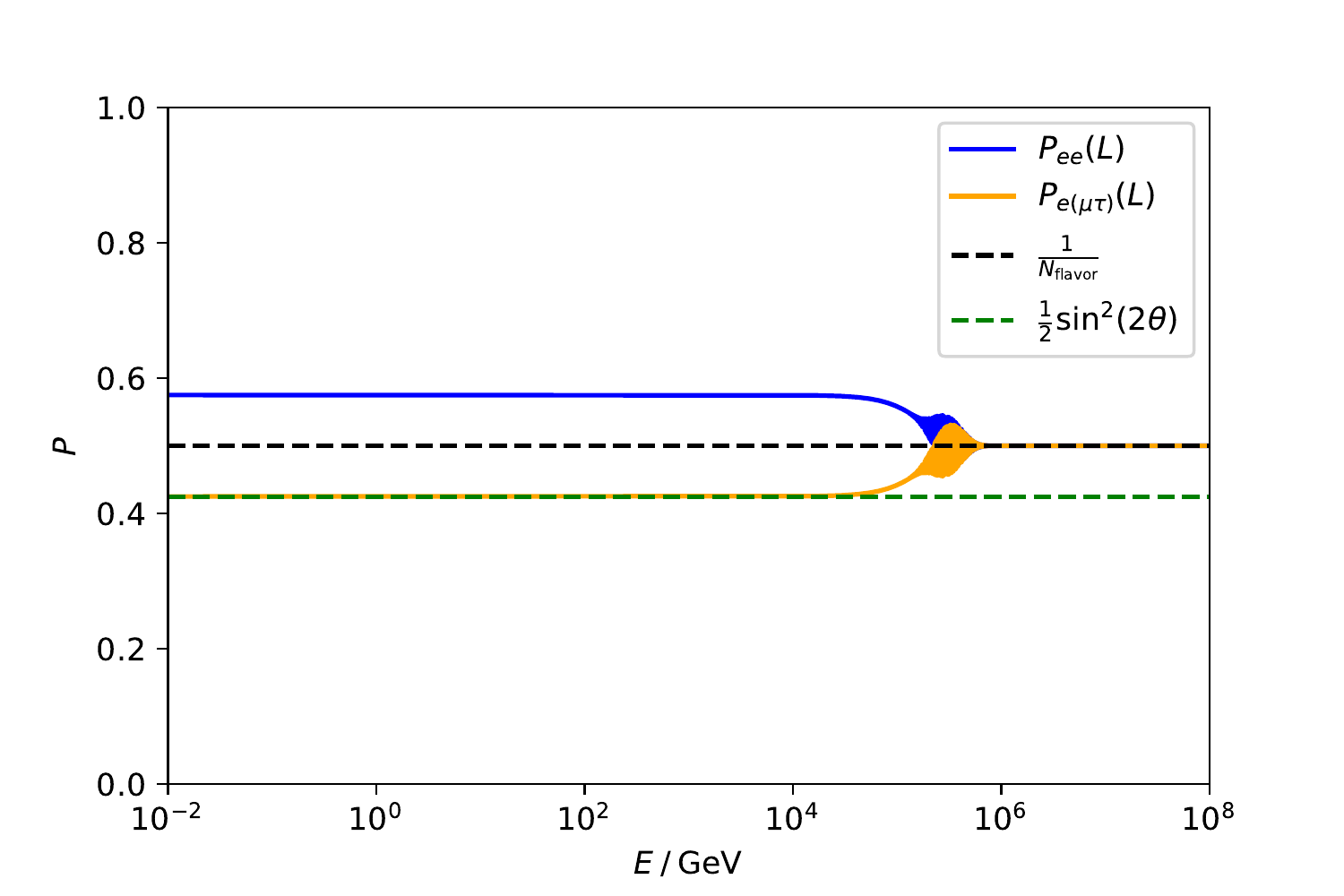}
    \caption{Energy dependence of the neutrino oscillation probabilities \(P_{ee}\) (blue) and \(P_{e(\mu\tau)}\) (orange) at a fixed baseline of \(L = 10\;\mathrm{kpc}\).
    The asymptotic values of \(\sfrac{1}{2}\sin^2(2\theta)\) and \(\sfrac{1}{N_{\mathrm{flavor}}}\) are shown as horizontal, dashed, green and black lines, respectively.}
    \label{fig:result_plot_E}
\end{figure}

More interesting than the baseline dependence is the energy dependence, though, displayed in
Fig.~\ref{fig:result_plot_E} at a fixed baseline of \(L = 10\;\mathrm{kpc}\)
 (the approximate distance to a potential neutrino source of electron antineutrinos in the direction of the Cygnus spiral arm~\cite{Anchordoqui:2005gj}).
The figure exhibits the fundamentally different energy behavior of the two types of decoherence considered:
While both types of decoherence become effective at large baselines, wave packet decoherence dominates low energy neutrinos (\(E \lesssim 100\;\mathrm{TeV}\)),
higher energy neutrinos still oscillate,
and neutrinos of the highest energies
\(E \gtrsim 1\;\mathrm{PeV}\) are subject to
quantum-gravitational decoherence that entails a rapid convergence towards a democratic flavor mix.
This prediction has to be compared with palatable flavor ratios obtained from wave packet separation
in the standard scenario that allows for
flavor compositions at Earth that range from
(0.6 : 1.3 : 1.1) to (1.6 : 0.6 : 0.8) \cite{Pakvasa:2007dc,Aartsen:2015ivb,Bustamante:2015waa}.
A serious obstacle for the discovery of quantum-gravitational decoherence is the fact that the canonical flavor ratio obtained from wave packet separation from a pion source is exactly a democratic 1:1:1 mix.
One possibility to discriminate quantum-gravity induced decoherence from wave packet decoherence is thus to identify point sources
with flavor ratios different from 1:1:1, originating e.g. from neutron decay. One candidate point source has been identified as Cygnus OB2 in the direction of the Cygnus spiral arm \cite{Anchordoqui:2005gj,Anchordoqui:2006pb}.
While
small statistics and limited angular resolution make the identification of neutrino point sources challenging, the KM3NET \cite{Huang:2020bdg} and GVD \cite{Stasielak:2021gcv} detectors under construction in the northern hemisphere will complement the IceCube sky coverage and the recent
IceCube-Gen2 proposal is designed to achieve an improved sensitivity to discriminate neutrino sources
\cite{Aartsen:2020fgd}.

\section{Decoherence in the presence of New Fermions}
\begin{figure}
    \centering
    \includegraphics[width=\textwidth]{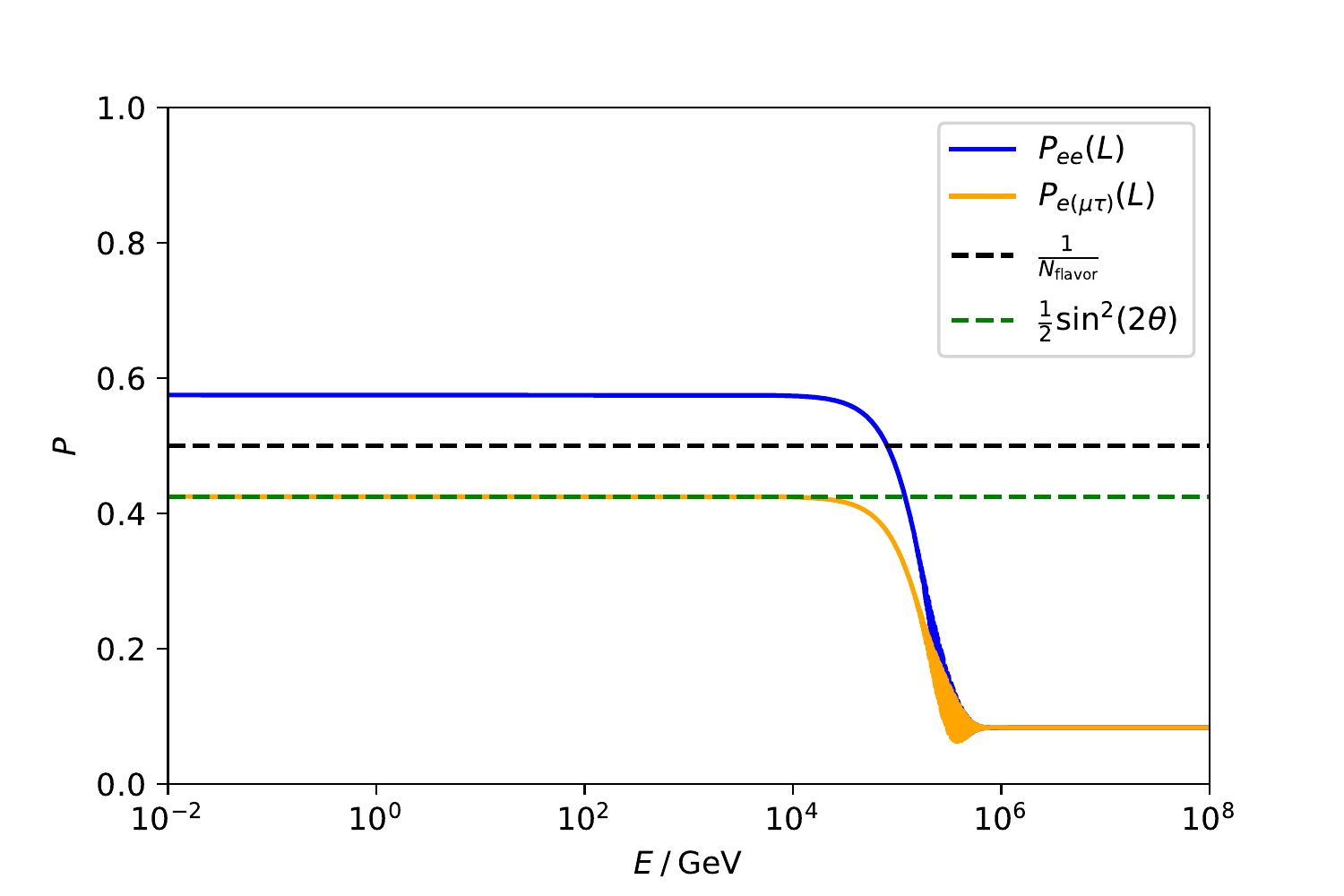}
    \caption{Energy dependence of the modified neutrino oscillation probabilities \(P_{ee}\) (blue) and \(P_{e(\mu\tau)}\) (orange) at a fixed baseline of \(L = 10\;\mathrm{kpc}\) with 10 additional neutral fermions in the model.
    The asymptotic values of \(\sfrac{1}{2}\sin^2(2\theta)\) and \(\sfrac{1}{N_{\mathrm{flavor}}}\) are shown as horizontal, dashed, green and black lines, respectively.}
    \label{fig:mod_result_plot_E}
\end{figure}

The problem to identify quantum-gravitational decoherence can be ameliorated by
a particularly interesting feature of the phenomenon under study that arises if there exist unknown neutral fermions in addition to the three known neutrino flavors. In this case,
quantum-gravitational effects are expected to cause a uniform distribution among all known and unknown fermions with equal (unbroken) gauge quantum numbers
after the beam has travelled a sufficiently large distance from the source to the detector.
Thus, considering a scenario with two Standard Model neutrinos and \(N - 2\) additional neutral fermion fields \(\chi_k\) not mixing with the neutrino sector, the oscillation formulae shown in Eq.~\eqref{eq:result1} and~\eqref{eq:result2} are altered in order to account for the loss of probability from the neutrino system to the newly introduced neutral flavors \(\chi_k\). Expanding the Lindblad equation and density matrix using the generators of \(SU(N)\) together with the identity matrix, we arrive at the modified oscillation probabilities
\begin{align}
    P_{ee}(L) &= \frac{1}{N} + \frac{N-2}{2N}e^{-2\gamma L} + \frac{1}{2}\cos^2(2\theta)e^{-2\gamma L}\nonumber\\
                &+\frac{1}{2}\sin^2(2\theta)e^{-\left(\gamma+\frac{1}{L_{\mathrm{coh}}}\right)L}\left\{\cos\left(\omega L\right)+ \frac{\gamma}{\omega}\sin\left(\omega L\right)\right\}\,,\label{eq:mod_result1}\\
    P_{e(\mu\tau)}(L)&= \frac{1}{N} + \frac{N-2}{2N}e^{-2\gamma L} - \frac{1}{2}\cos^2(2\theta)e^{-2\gamma L}\nonumber\\
                &-\frac{1}{2}\sin^2(2\theta)e^{-\left(\gamma+\frac{1}{L_{\mathrm{coh}}}\right)L}\left\{\cos\left(\omega L\right)+ \frac{\gamma}{\omega}\sin\left(\omega L\right)\right\}\, \label{eq:mod_result2}\,.
\end{align}
For a proof of these formulae see Appendix~\ref{app:proof}.
Again, for the case where \(\exp(-2\gamma L) \approx 1\) the \(2\nu\) oscillation formula with wave packet decoherence is recovered.
This is expected because the neutrino and \(\chi\) sectors are completely decoupled from each other as long as quantum-gravitational decoherence effects are negligible.
For much larger baselengths or energies the exponential factor decays to \(0\) and leaves the asymptotic \(\sfrac{1}{N}\) behavior entailing a uniform flavor mix.
This behavior is highlighted in Fig.~\ref{fig:mod_result_plot_E} where we plot the energy dependence
at a fixed baseline \(L = 10\;\mathrm{kpc}\) for a hypothetical scenario of \(N - 2 = 10\) additional neutral fermions.
As can be seen,
Eqs.~\eqref{eq:mod_result1} and~\eqref{eq:mod_result2} exhibit the same oscillation behavior as Eqs.~\eqref{eq:result1} and~\eqref{eq:result2} in the regime where quantum-gravity effects are negligible, but exhibit a very different behavior, i.e. drop to \(\sfrac{1}{12}\) instead of \(\sfrac{1}{2}\), at very high energies where these effects become significant.
\FloatBarrier 

A generalization to the full $3\nu$ formalism is straightforward but lengthy, and will be presented elsewhere. We thus confine ourselves here to point out that the asymptotic behavior in the $3\nu$ case can be easily inferred by comparing the $2\nu$ formalism
discussed here with the $3\nu$ formalism presented e.g. in~\cite{Gago:2002na,Barenboim:2006xt}, cf. eqs. (21) and (2.19) in these references, respectively. In both cases the high energy asymptotic approaches a democratic mix of all $N$ electrically neutral quantum fields, so that the proposed signature of a drop in the total flux from $1$ to $2/N$ deduced here is simply replaced by the corresponding asymptotic of $3/N$ above the critical threshold. The resulting asymptotic oscillation probability will amount to 1/3 instead of 1/2 in Figs.~1,2 and 1/13 instead of 1/12 in Fig.~3.

\section{Discussion and Conclusion}
\begin{figure}
    \centering
    \includegraphics[width=\textwidth]{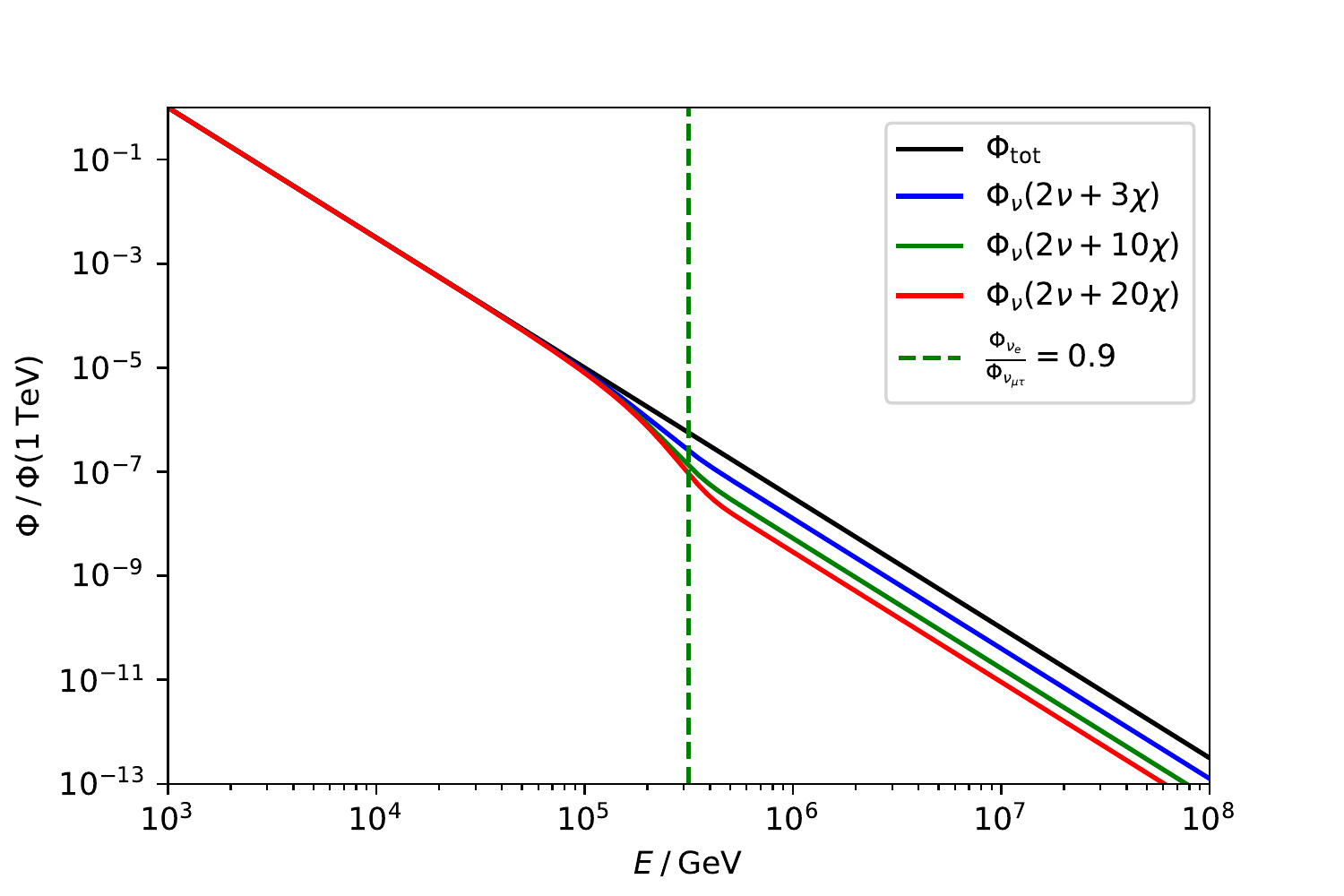}
    \caption{Energy dependence of the total flux (\(\Phi_{\mathrm{total}}\)) of neutral particles in black compared to the summed neutrino fluxes \(\Phi_{\nu}\) for 3 (blue), 10 (green) and 20 (red) additional neutral fermions in the model. A dip in the \(\Phi_\nu\) spectrum occurs after approximate flavor equilibration \(\sfrac{\Phi_{\nu_e}}{\Phi_{\nu_{\mu\tau}}} = 0.9\) is reached.
    The corresponding energy of flavor equilibrium is indicated by the green, dashed, vertical line. Afterwards the flavor ratio converges to 1.}
    \label{fig:flux_dip}
\end{figure}

Thus, if there exist undiscovered neutral fermions not included in the Standard Model, as a consequence of quantum-gravitational-decoherence equilibrating
an original flux of astrophysical neutrinos over all flavors we expect a dip
in the total neutrino flux setting in at the threshold energy of the quantum-gravitational effect.
The position of the dip will coincide with the transition to a democratic flavor mix in cases where the flavor mix
resulting from wave packet decoherence at low energies is non-democratic.
This effect depends on the number of neutral degrees of freedom and can range from 25\% for a single, additional sterile neutrino or WIMP up to a dramatic cutoff in models with a large number of new particles,
e.g.
Kaluza-Klein excitations in models with extra space dimensions \cite{ArkaniHamed:1998rs}
or scenarios with a large number of
Standard Model copies \cite{Dvali:2007hz}.
In Fig.~\ref{fig:flux_dip}, we show the total flux \(\Phi_{\mathrm{total}}\) of neutral fermions stemming from an astrophysical neutrino source with \(\Phi_{\mathrm{total}}(E)\propto E^{-2.5}\) and the corresponding total neutrino fluxes after a travel distance of \(L = 10\;\mathrm{kpc}\) at earth for different numbers of additional neutral fermions.

In this context it is interesting to observe that IceCube so far hasn’t observed any astrophysical neutrinos above
10~PeV. This has inspired speculations
about a possible break in the astrophysical neutrino spectrum \cite{Anchordoqui:2014hua,Learned:2014vya,Palomares-Ruiz:2015mka,Mohanty:2018cmq}
although the detection of an event attributed to the Glashow resonance at 6.3~PeV has been reported recently \cite{IceCube:2021rpz}.
While it has been pointed out that antineutrinos may feature different flavor ratios as compared to neutrinos and that this fact may make it
difficult to pinpoint the original flavor ratios at the source \cite{Nunokawa:2016pop}, in our case this effect is beneficial rather than problematic as
quantum-gravity induced decoherence will equilibrate the original astrophysical neutrino flux over all neutral degrees of freedom including
antiparticles so that a transition from non-democratic flavor ratios to democratic ones at a certain energy threshold will provide an indication
for quantum gravity irrespective of the original spectra.

Note that the dip predicted here is expected to be smoothed by suppression factors of powers
$\sim(m/E)^{2k}$
for particles of mass \(m\) having a weak isospin or other hypothetical broken gauge quantum numbers different from neutrinos.
Here \(E\) is the neutrino beam energy and $k$ is the number of necessary insertions of the respective Higgs vacuum expectation values.
It is of course also smeared out by traveling neutrons and backgrounds from the diffuse flux from other sources.
Finally, features in the astrophysical neutrino spectra may also have their origin in the production mechanisms.

\FloatBarrier 

The scenario discussed in this paper is speculative in several respects. On the one hand, it is model-dependent how quantum spacetime interacts with particles propagating in vacuo, whether this interaction indeed breaks global quantum numbers and whether these effects are large enough to be probed experimentally. On the other hand, it is not clear when neutrino telescopes will be capable to identify galactic or even extragalactic point sources that produce neutrinos via neutron decays and accumulate sufficient statistics
to test the scenario discussed here. We nevertheless feel that a thorough study of the effect discussed is appropriate, for the following reasons. First,
the breaking of global quantum numbers and the consequential equilibration of original neutrino fluxes over all degrees of freedom sharing the same gauge quantum numbers seems to be a rather generic prediction of quantum gravity and is the only crucial assumption made here. Thus if this expectation wouldn't be
realized in quantum-gravity induced decoherence that would be an interesting result by itself and may shed some light on the highly relevant question
about how black holes process information. Next, first point sources of high-energy astrophysical neutrino fluxes have been identified already in the IceCube experiment  \cite{IceCube:2018cha,Abbasi:2021bvk}. As has been argued in \cite{Anchordoqui:2006pb}, multi-messenger astronomy can identify the neutrino flux accompanying cosmic ray acceleration in the Cygnus spiral arm with an evidence at the $5~\sigma$ level in 15 years of observation. Finally, the potential relevance of the effect discussed here can hardly be overestimated, as it provides a unique window into hidden sectors and thus one of the most pressing problems in present-day particle physics and cosmology.

\section*{Acknowledgements}
We thank Luis Anchordoqui, John Learned, Sergio
Palomares-Ruiz and Tom Weiler for helpful discussions. ER is supported by The 5000 Doktor scholarship Program by
The Ministry of Religious Affairs, Indonesia.

\newpage
\appendix
\section{Proof of the N Level Oscillation Formulae}
\label{app:proof}
Now, we want to proof the expression for the neutrino oscillation probabilities~\eqref{eq:mod_result1} and~\eqref{eq:mod_result2} if \(N-2\) (\(N \geq 2\)) additional neutral fermions are present in the model.
In order to proof the respective formulae, we are using the density matrix formalism in the mass eigenstate basis of the system and employ the following assumptions:
\begin{enumerate}
    \item The system propagates in vacuum with approximately the same momentum \(p\) for all mass eigenstates
    \item The neutrino mass is much smaller than the momentum \(p \gg m\), hence we use \(E \approx p\) in terms of \(\mathcal{O}(m_{\nu})\)
    \item The additional neutral fermions are not mixed with the (active) neutrino sector
    \item The mixing of the 2 neutrinos is fully characterized by the mixing angle \(\theta\) and the mass squared difference \(\Delta m^2\)
    \item Decoherence induced by wave paket separation leads to an exponential dampening of the off diagonal elements of the density matrix \(\varrho\) according to the coherence length~\cite{Kersten:2015kio}
    \begin{align}
        L_{\mathrm{wp}} := \frac{\sigma_x}{\Delta v_{ij}} = \frac{\sigma_x E}{\Delta E_{ij}}
    \end{align}
    where \(\sigma_x\) is the initial wave paket size and \(\Delta E_{ij} = E_i - E_j\) is the difference of the Hamiltonian eigenvalues \(E_i\) and \(E_j\)
    \item Decoherence induced by quantum gravity is caused by interactions of the system with the hypothetical spacetime foam during which a certain mass eigenstate is selected democratically.
    Furthermore, quantum gravitationally induced decoherence is described by only one parameter \(\Gamma(E)\), as described in~\cite{Stuttard:2020qfv}.
\end{enumerate}
Using these prerequisites we arrive at the oscillation probabilities
\begin{align}
    P_{ee}(L) &= \frac{1}{N} + \frac{N-2}{2N}e^{-2\gamma L} + \frac{1}{2}\cos^2(2\theta)e^{-2\gamma L}\nonumber\\
                &+\frac{1}{2}\sin^2(2\theta)e^{-\left(\gamma+\frac{1}{L_{\mathrm{coh}}}\right)L}\left\{\cos\left(\omega L\right)+ \frac{\gamma}{\omega}\sin\left(\omega L\right)\right\}\,,\\
    P_{e(\mu\tau)}(L)&= \frac{1}{N} + \frac{N-2}{2N}e^{-2\gamma L} - \frac{1}{2}\cos^2(2\theta)e^{-2\gamma L}\nonumber\\
                &-\frac{1}{2}\sin^2(2\theta)e^{-\left(\gamma+\frac{1}{L_{\mathrm{coh}}}\right)L}\left\{\cos\left(\omega L\right)+ \frac{\gamma}{\omega}\sin\left(\omega L\right)\right\}\,,
\end{align}
where \(\gamma = \sfrac{\Gamma}{2}\).
\newpage
\textbf{Proof:}\\
In order to prove the oscillation formulae shown in Eqs.~\eqref{eq:mod_result1} and~\eqref{eq:mod_result2}, we first derive the Lindblad equation for the \(2\nu\) + \((N - 2)\chi\) system.
This equation is then expressed in a specific basis \(\mathcal{B}\) of the vector space of \(N \times N\) hermitian matrices, \(\mathbb{H}(N)\), namely the \(SU(N)\) equivalents of the Pauli matrices.
These matrices can be split into the set of diagonal and off-diagonal matrices and we adopt the following basis ordering
\begin{align}
    \mathcal{B} &= \{\underbrace{\lambda_0}_{\propto \mathbbm{1}},\underbrace{\lambda_1,\ldots,\lambda_{N(N-1)}}_{\mathrm{off-diagonal}},\underbrace{\lambda_{N(N-1)+1},\lambda_{N^2-1}}_{\mathrm{diagonal}}\}\,.
\end{align}
Here, we implicitly use the fact that there are exactly \(N(N-1)\) off-diagonal Pauli-like matrices and \((N-1)\) diagonal ones.
In order to get a complete basis, we need to include a matrix proportional to the identity, i.e. \(\lambda_0\).

\subsection{SU(N) Pauli-like Matrices}
\label{ssec:Pauli}
Now, we discuss the explicit shape and properties of the aforementioned basis matrices.

\subsubsection{Off-diagonal Pauli-like Matrices}
\label{sssec:off-diag}
In the following, the off-diagonal \(\lambda_k\) are of major importance, hence we want to discuss them first.
In general, they are very simple in shape since they only contain two nonzero entries at indices \((j_0, k_0)\) with \(k_0 > j_0\) and the respective transposed position \((k_0, j_0)\) where the complex conjugate of the \((j_0, k_0)\) element is filled in such that they become hermitian matrices.
Moreover, these nonzero entries alternate between \(1\) and \(i\).
In total, this results in a simple formula for the \((a,b)\) element of a certain off-diagonal \(\lambda_k\), i.e.
\begin{align}
    [\lambda_{\mathrm{off}}]_{ab} = \zeta^{\ast} \delta_{j_0 a}\delta_{k_0 b} + \zeta \delta_{k_0 a}\delta_{j_0 b}\,,
\end{align}
with \(\zeta\) alternating between \(1\) and \(i\) for neighboring matrices.
This special shape of the off-diagonal matrices has an important consequence:
Let \(A \in \mathbb{H}(N)\), then \(A = A^\dagger\) by definition and therefore each off-diagonal \(a_{jk}\) element of \(A\) in its lower triangle (\(j > k\)) is related to the off-diagonal element \(a_{kj} = a_{jk}^{\ast}\) in its upper triangle.
In the following, we call this a \textit{related pair} of off-diagonal elements and define the real (imaginary) part of the related pair as the real (imaginary) part of the element in the lower triangle.\\
Using this definition, we see that each off-diagonal \(\lambda_k\) directly corresponds to the real or imaginary part of such a related pair.
Here, correspondence means that the component \(\hat{a}_k := \langle A, \lambda_k \rangle\) of \(A\) with respect to \(\mathcal{B}\), where \(\lambda_k\) is off-diagonal, equals to the real (\(\zeta = 1\)) or imaginary part (\(\zeta = i\)) of a related pair of off-diagonal elements.
This correspondence is one-to-one and is a key point of the following discussion.
Schematically, we can express this the following way
\begin{align}
    A = \begin{pmatrix}
        \ast & \ast         & \ast          & \ast          & \ast & \ast \\
        \ast & \ast         & \ast          & \ast          & \ast & \ast \\
        \ast & \ast         & \ast          & a_{j_0 k_0}   & \ast & \ast \\
        \ast & \ast         & a_{k_0 j_0}   & \ast          & \ast & \ast \\
        \ast & \ast         & \ast          & \ast          & \ast & \ast \\
        \ast & \ast         & \ast          & \ast          & \ast & \ast \\
    \end{pmatrix} \sim \underbrace{\mathrm{Re}(a_{j_0 k_0})}_{\hat{a}_k} \lambda_{k} + \underbrace{\mathrm{Im}(a_{j_0 k_0})}_{\hat{a}_{k+1}}\lambda_{k+1}\,,
\end{align}
where \(\sim\) means that the explicitly mentioned related pair \((a_{j_0 k_0}, a_{k_0 j_0})\) on the left hand side is put into the right position by evaluating the right hand side.\\
Lastly, if we want to perform explicit calculations using this basis, we need to specifiy the ordering of these off-diagonal matrices.
For simplicity, we use the same ordering scheme as is used for the Gell-Mann matrices, i.e. neighboring \(\lambda_k\) alternate between \(\zeta = 1\) and \(\zeta = i\), as already mentioned, and the indices \((j_0,k_0)\) of nonzero \(\lambda_k\)-elements are ordered such that for the first \(N-1\) matrix pairs \(j_0 = 1\) and \(k_0\) runs between \(2\) and \(N\), for the next \(N-2\) pairs we then choose \(j_0 = 2\) and \(k_0\) runs between \(3\) and \(N\) and so on. 
For e.g. \(N = 3\) this resuts in
\begin{align}
    \lambda_1 &= \begin{pmatrix}
        0 & 1 & 0\\
        1 & 0 & 0\\
        0 & 0 & 0\\
    \end{pmatrix}\,,\,\,
    \lambda_2 = \begin{pmatrix}
        0 & -i & 0\\
        i & 0 & 0\\
        0 & 0 & 0\\
    \end{pmatrix}\,,\,\,
    \lambda_3 = \begin{pmatrix}
        0 & 0 & 1\\
        0 & 0 & 0\\
        1 & 0 & 0\\
    \end{pmatrix}\,,\\
    \lambda_4 &= \begin{pmatrix}
        0 & 0 & -i\\
        0 & 0 & 0\\
        i & 0 & 0\\
    \end{pmatrix}\,,\,\,
    \lambda_5 = \begin{pmatrix}
        0 & 0 & 0\\
        0 & 0 & 1\\
        0 & 1 & 0\\
    \end{pmatrix}\,,\,\,
    \lambda_6 = \begin{pmatrix}
        0 & 0 & 0\\
        0 & 0 & -i\\
        0 & i & 0\\
    \end{pmatrix}\,.
\end{align}

\subsubsection{Diagonal Basis Matrices}
\label{sssec:diag}
After having discussed the off-diagonal \(\lambda_k\), we now turn towards the (simpler) diagonal ones.
These are comprised of the \(N-1\) diagonal Pauli-like matrices
\begin{align}
    \lambda_{N(N-1) + m} = \sqrt{\frac{2}{m(m+1)}}\mathrm{diag}(\underbrace{1, \ldots, 1}_{m\times}, -m, 0, \ldots, 0)\,,\,\,1 \leq m \leq N-1\,,
\end{align}
and the scaled identity
\begin{align}
    \lambda_0 := \sqrt{\frac{2}{N}} \mathbbm{1}\,.
\end{align}
As indicated above, the number of ones on the diagonal of the diagonal Pauli-like matrices is exactly \(m\) such that these matrices are traceless.
For \(N=3\) this results in
\begin{align}
    \lambda_0 &= \sqrt{\frac{2}{3}}\begin{pmatrix}
        1 & 0 & 0\\
        0 & 1 & 0\\
        0 & 0 & 1\\
    \end{pmatrix}\,,\quad
    \lambda_7 &= \begin{pmatrix}
        1 & 0 & 0\\
        0 & -1 & 0\\
        0 & 0 & 0\\
    \end{pmatrix}\,,\quad
    \lambda_8 = \sqrt{\frac{1}{3}}\begin{pmatrix}
        1 & 0 & 0\\
        0 & 1 & 0\\
        0 & 0 & -2\\
    \end{pmatrix}\,.
\end{align}
These matrices, together with the off-diagonal \(\lambda_k\), exactly resemble the Gell-Mann matrices of \(SU(3)\) with the ordering imposed as above.

\subsubsection{Properties of the Basis Matrices}
\label{sssec:prop}
The basis \(\mathcal{B}\) forms an orthonormal basis with respect to the scalar product
\begin{align}
    \langle \cdot, \cdot\rangle &: \mathbb{H}(N) \times \mathbb{H}(N) \rightarrow \mathbb{R}\\
    \langle A,B \rangle &:= \frac{1}{2}\mathrm{Tr}\left(A \cdot B \right)\,.
\end{align}
Moreover, \(\mathcal{B}\) fulfills the completenes relation
\begin{align}
    \mathbbm{I}[\cdot] &= \sum_{k = 0}^{N^2-1}\lambda_k \bar{\lambda}_k[\cdot]\,,\\
    \bar{\lambda}_k[A] &:= \frac{1}{2}\mathrm{Tr}(\lambda_k A)\quad \text{with} \quad A \in \mathbb{H}(N)\,,
\end{align}
where we introduced the identity operator \(\mathbb{I}\) over \(\mathbb{H}(N)\) and the dual basis \(\bar{\lambda}_k \in \mathcal{B}^\ast \subset \mathbb{H}(N)^\ast\).
In the following, we will partially employ the standard bra-ket notation for vectors in \(\mathbb{H}(N)\) where it is beneficial.

\subsection{Lindblad Equation of the System}
\label{ssec:Lindblad}
Now that we have specified the basis \(\mathcal{B}\) of matrices we are working in, we proceed by expanding the density matrix \(\varrho\) and all operators from the Lindblad equation in it.
The Lindlblad equation reads
\begin{align}
    \dot{\varrho} &= -i[H,\varrho] + \mathcal{D}_{\mathrm{wp}}[\varrho] + \mathcal{D}_{\mathrm{qg}}[\varrho] := \mathcal{L}[\varrho] \,, \label{eq:Lindblad_app}
\end{align}
where \(H\) is the Hamiltonian of the neutral fermion system and \(\mathcal{D}_{\mathrm{wp}}\) and \(\mathcal{D}_{\mathrm{qg}}\) are the dissipators describing wave paket and quantum gravitationally induced decoherence, respectively.
Furthermore, we define the full Lindblad operator \(\mathcal{L}\).\\
In order to differentiate between vectors in \(\mathbb{H}(N)\) and their components with repect to the basis \(\mathcal{B}\), we use curly symbols for the vectors and straight symbols for the components.
For example, the density matrix is denoted by \(\varrho\), while its components are denoted by \(\rho_k\) such that
\begin{align}
    \varrho = \sum_{k = 0}^{N^2-1} \rho_k \lambda_k\,.
\end{align}
The same notation is employed for operators acting on \(\mathbb{H}(N)\), for example \(\mathcal{D}_{\mathrm{wp}}\), and their matrix elements with respect to \(\mathcal{B}\), for example \(D_{\mathrm{wp}}\).

\subsubsection{Commutator}
\label{sssec:comm}
First, we start with the derivation of the general representation matrix \(C\) of the commutator part of the Lindblad equation, i.e.
\begin{align}
    \mathcal{C}[\varrho] := -i\left[H,\varrho\right]\,.
\end{align}
Here, \(H\) represents the vacuum Hamiltonian reading
\begin{align}
    H &= \mathrm{diag}\left(E_{\nu_1}, E_{\nu_2},E_{\chi_1},\ldots,E_{\chi_{N-2}}\right)\\
    &:= \mathrm{diag}\left(E_{1}, E_{2},E_{3},\ldots,E_{N}\right)\\
\end{align}
in the mass eigenstate basis of the system.
For this Hamiltonian, the components of the commutator read
\begin{align}
    (\mathcal{C}[\varrho])_{jk} &= -i([H,\varrho])_{jk} \\
    &= -i \sum_{l = 1}^{N} H_{jl}\varrho_{lk} - \varrho_{jl}H_{lk} \\
    &= -i \sum_{l = 1}^{N} E_j \delta_{jl} \varrho_{lk} - \varrho_{jl} E_l \delta_{lk} \\
    &= -i \left(E_j - E_k\right) \varrho_{jk} \\
    &:= \Delta E_{jk} (\mathrm{Im}(\varrho_{jk}) - i \mathrm{Re}(\varrho_{jk}))\\
    &= -\Delta E_{jk} \rho_{l+1} - i \Delta E_{jk}\rho_{l}\\
    &= \langle \mathcal{C}[\varrho], \lambda_{l}\rangle - i \langle \mathcal{C}[\varrho], \lambda_{l+1} \rangle\,,
\end{align}
where we substituted in the coefficients of \(\varrho\) and \(\mathcal{C}[\varrho]\) in the basis \(\mathcal{B}\) (the coefficient index \(l\) depends on the matrix indices \((j,k)\)) in the last two lines in order to be able to directly read off the matrix elements.\\
Here we need to be careful because depending on if we consider the case \(j<k\) or \(j>k\) the imaginary part of \(\varrho_{jk}\) corresponds to either \(-\rho_{l+1}\) (\(j<k\)) or \(+\rho_{l+1}\) (\(j>k\)) and the same holds for \(C[\varrho]\) and its coefficients, because both matrices are hermitian.
Independent of the choice we will obtain the same matrix element, which is why we chose \(j < k\) for convenience.
In order to derive the representation matrix \(C\) of the operator \(\mathcal{C}[\cdot] = -i[H,\cdot]\), we have to make some crucial observations from the previous equations:
\begin{enumerate}
    \item Diagonal elements of the resulting matrix \((\mathcal{C}[\varrho])_{jj}\) vanish, since \(\Delta E_{jj} = 0\).
    \item \((\mathcal{C}[\varrho])_{jk} \propto \varrho_{jk}\), no other components of \(\varrho\) contribute to \((\mathcal{C}[\varrho])_{jk}\).
    \item The coefficient \(\rho_{l+1} = \mathrm{Im}(\varrho_{kj}))\) is scaled by \(-\Delta E_{jk}\) and becomes the real part of \((\mathcal{C}[\varrho])_{jk} = \langle \mathcal{C}[\varrho], \lambda_{l}\rangle\).
    \item The coefficient \(\rho_{l} = \mathrm{Re}(\varrho_{kj}))\) is scaled by \(\Delta E_{jk}\) and then becomes the imaginary part of \((\mathcal{C}[\varrho])_{jk} = \langle \mathcal{C}[\varrho], \lambda_{l+1}\rangle\).
\end{enumerate}
These 4 properties of the result of \(\mathcal{C}\) operating on an \textit{arbitrary} hermitian matrix \(\varrho\) completely determine the form of its representation matrix \(C\) in the basis \(\mathcal{B}\).\\
To clarify our explanations below, we shortly recall the meaning of the matrix elements of \(C = (C_{ij})_{i,j = 0}^{N^2-1}\).
If we expand \(\mathcal{C}\) in the basis \(\mathcal{B}\), we obtain
\begin{align}
    \mathcal{C}\left[\varrho\right] &= \mathcal{C} \vert \varrho \rangle = \sum_{j = 0}^{N^2 - 1} \rho_j \mathcal{C}\vert \lambda_j \rangle \\
    &= \sum_{j = 0}^{N^2 - 1} \rho_j \mathbb{I}\mathcal{C}\vert \lambda_j \rangle \\
    &= \sum_{j = 0}^{N^2 - 1} \sum_{i = 0}^{N^2 - 1} \rho_j \vert \lambda_i \rangle \underbrace{\langle \lambda_i \vert \mathcal{C} \vert \lambda_j \rangle}_{= C_{ij}}\\
    &= \sum_{i = 0}^{N^2 - 1} \left(\sum_{j = 0}^{N^2 - 1} C_{ij} \rho_j\right)  \vert \lambda_i \rangle\,.
\end{align}
This means that the elements \(C_{ij}\) for fixed \(i\) and \(j\) determine the contribution of the \(j\)-th component \(\rho_j\) of \(\varrho\) to the \(i\)-th component of the result.
Using this and the first observation from the enumeration above, we know that all rows \(i\) of \(C\) corresponding to diagonal basis matrices have to vanish.
Otherwise one could always find a matrix \(\varrho \in \mathbb{H}(N)\), such that \(\mathcal{C}[\varrho]\) has diagonal entries, which contradicts our findings from above.
Equally, the result \(\mathcal{C}[\varrho]\) cannot contain \(\rho_i\) components contributing to the diagonal of \(\varrho\) due to observations number 1 and 2.
Hence, we find
\begin{align*}
    \boxed{C_{ij} = C_{ji} = 0\,, \quad \forall i = 0, N(N-1) + 1, \ldots, N^2-1\,,\forall j = 0, \ldots, N^2-1}\,.
\end{align*}
Furthermore, the one-to-one correspondence of \(\lambda_k\) to related pairs of off-diagonal elements of \(\varrho\) together with observation number 2 implies that \(C\) is a block diagonal matrix, since only the real and imaginary parts of \(\varrho_{jk}\) contribute to the real and imaginary parts of the resulting matrix element \((\mathcal{C}[\varrho])_{jk}\).
Therefore, we arrive at the following form of \(C\)
\begin{align}
    C &= \begin{pmatrix}
        0       &               & \vec{0}^T                 \\
                & \tau   & \underline{0}^T           \\
        \vec{0} & \underline{0} & \underline{\underline{0}} \\
    \end{pmatrix}\in \mathbb{R}^{N^2 \times N^2}\\
    \tau &= \begin{pmatrix}
        \ast    & \ast      &           &           &       \\
        \ast    & \ast      &           &           &       \\
                &           & \ddots    &           &       \\
                &           &           & \ast      & \ast  \\
                &           &           & \ast      & \ast  \\
    \end{pmatrix} \in \mathbb{R}^{N(N-1) \times N(N-1)} \,,
\end{align}
where \(\vec{0} \in \mathbb{R}^{N^2-1}\), \(\underline{0} \in \mathbb{R}^{(N-1) \times N(N-1)}\) and \(\underline{\underline{0}} \in \mathbb{R}^{(N-1) \times (N-1)}\).
Finally according to observations 3 and 4, we know that only the real part of \(\varrho_{jk}\) contributes to the imaginary part of \((\mathcal{C}[\varrho])_{jk}\), while only the imaginary part of \(\varrho_{jk}\) contributes to the real part of \((\mathcal{C}[\varrho])_{jk}\).
The coefficients these parts are scaled with determine the corresponding matrix element of \(C\).
Hence, we arrive at its final form, i.e.
\begin{align}
    C &= \mathrm{Bdiag}\left(0, A_{12}, A_{13}, \ldots, A_{(N-1)N}, \underline{\underline{0}}\;\right)\,,\\
    A_{ij} &= \begin{pmatrix}
        0               & -\Delta E_{ij} \\
        \Delta E_{ij}   & 0             \\
    \end{pmatrix}\,.
\end{align}
Here, the \(\mathrm{Bdiag}\) instruction yields a block diagonal matrix comprised of the corresponding matrices given in its argument list.
The ordering of \(A_{ij}\) matrices in this construction resembles the same ordering scheme we use for the off-diagonal \(\lambda_k\), i.e. for each fixed \(i\) we iterate over all possible \(j\) values, greater than \(i\), and after one full \(j\)-iteration we increase \(i\).

\subsubsection{Dissipator}
\label{sssec:dis}
Next, we consider the dissipator \(\mathcal{D}\) introducing the open system effects into the time evolution of the density matrix.
As discribed in the prerequisites, we focus on two kinds of decoherence, i.e. wave paket separation and quantum gravitational decoherence.\\
First, we determine the wave packet decoherence influence.
For each pair of fermions, neutrinos obviously included, we have a respective coherence length determining the strength of coherence dampening, based on the differences of the eigenvalues of the Hamiltonian.
This coherence dampening translates to dampening of off-diagonal \(\varrho\) elements, since these describe the coherence of different mass eigenstates.
Assuming that the real and imaginary parts of the off-diagonal elements are damped with the same strength and that these off-diagonal elements \(\varrho_{jk}\) are only affected by the respective energy splitting \(\Delta E_{jk}\), we find that \(D_{\mathrm{wp}}\) is fully diagonal in the basis \(\mathcal{B}\).
It reads
\begin{align}
    D_{\mathrm{wp}} &= \mathrm{diag}\left(0,\frac{-1}{L_{12}},\frac{-1}{L_{12}},\frac{-1}{L_{13}},\frac{-1}{L_{13}},\ldots,\frac{-1}{L_{(N-1)N}}, \frac{-1}{L_{(N-1)N}},\underbrace{0,\ldots,0}_{N-1\text{ times}}\right)\,.
\end{align}
Lastly considering quantum gravitationally induced decoherence, we assume, in accordance with~\cite{Stuttard:2020qfv}, a democratic selection of mass eigenstates during the propagation of the beam.
This leads to averaging over all neutral flavor components of the beam, since each superposition of mass eigenstates is projected onto one certain mass eigenstate with equal probability.
This effect is simplest described by a dissipator of the form
\begin{align}
    D_{\mathrm{qg}} &= \mathrm{diag}\left(0,-\Gamma_1,\ldots,-\Gamma_{N^2-1}\right)\,.
\end{align}
The zeroth entry of the diagonal has to vanish in order to conserve the overall probability of the system, which is important, since in our scenario no probability is lost to the environment.
Furthermore, the entries \(\Gamma_{1},\ldots, \Gamma_{N(N-1)}\) describe coherence dampening of the mass eigenstates due to the projection onto certain mass eigenstates and have essentially the same effect as the wave paket decoherence mechanism.
As in Ref.~\cite{Liu:1997zd} we choose \(\Gamma_1 = 0\).
Lastly, the most interesting feature of quantum gravitationally induced decoherence is encoded in the last \(N-1\) entries corresponding to the diagonal \(\lambda_k\) matrices.
This means that after a sufficiently long travel distance only the part of \(\varrho\) proportional to the identity remains undamped.
Hence, these entries lead to the dampening of any excess number of mass eigenstates in the beam and therefore describe the democracy in the selection of mass eigenstates.
Following again Ref.~\cite{Stuttard:2020qfv}, we choose \(\Gamma = \Gamma_k\) for all \(k = 2,\ldots, N^2-1\).\\
Now we are able to assemble the representation matrix \(\Lambda\) of the full Lindblad operator \(\mathcal{L}\), i.e.
\begin{align}
    \Lambda &:= C + D_{\mathrm{wp}} + D_{\mathrm{qg}}\\
    \Lambda &= \mathrm{Bdiag}\left(0, \Lambda_{12}^{\nu}, \Lambda_{13}, \ldots, \Lambda_{(N-1)N}, -\Gamma,\ldots,-\Gamma\right)\\
    \Lambda_{ij}^{\nu} &= \begin{pmatrix}
        -\frac{1}{L_{ij}} & -\Delta E_{ij}\\
        \Delta E_{ij} & -\frac{1}{L_{ij}} - \Gamma
    \end{pmatrix}\,,\quad\Lambda_{ij} = \begin{pmatrix}
        -\frac{1}{L_{ij}}-\Gamma & -\Delta E_{ij}\\
        \Delta E_{ij} & -\frac{1}{L_{ij}} - \Gamma
    \end{pmatrix}\,.
\end{align}
Due to the special block diagonal form of the Lindblad operator in vacuum, we can solve the system of equations analytically for arbitrary \(N \geq 2\) by employing diagonalizing matrices of the form
\begin{align}
    S &= \mathrm{Bdiag}\left(1, S_{12}^{\nu}, S_{13}, \ldots, S_{(N-1)N},\mathbbm{1}_{(N-1)\times (N-1)}\right)\\
    S^{-1} &= \mathrm{Bdiag}\left(1, {S_{12}^{\nu}}^{-1}, S_{13}^{-1}, \ldots, S_{(N-1)N}^{-1},\mathbbm{1}_{(N-1)\times (N-1)}\right)\,,
\end{align}
such that \(\Lambda = S \tilde{D} S^{-1}\) and each \(S_{ij}\) diagonalizes the corresponding \(\Lambda_{ij} = S_{ij}\tilde{D}_{ij}S_{ij}^{-1}\).
This form of the Lindblad equation is also valid for more than two neutrino generations, since until now the framework is independent of the special type of neutral fermions included.
This changes as soon as we specify the mixing of flavor eigenstates of the fermions.

\subsection{Oscillation Probabilities}
Now, again considering the \(2\nu\) + \((N-2)\chi\) case, we are especially interested in the scenario where we start with an initially pure neutrino state.
This is needed in order to derive the neutrino oscillation probabilities from one neutrino flavor into another (not necessarily different) flavor.
For example, starting with a pure electron flavor at the source the respective initial density matrix reads
\begin{align}
    \varrho_{e} &= \begin{pmatrix}
        \cos^2(\theta)              & \cos(\theta)\sin(2\theta)     & 0         & \cdots & 0      \\
        \cos(\theta)\sin(2\theta)   & \sin^2(\theta)                & 0         & \cdots & 0      \\
        0                           & 0                             & 0         & \cdots & 0      \\
        \vdots                      & \vdots                        & \vdots    & \cdots & \vdots \\
        0                           & 0                             & 0         & 0      & 0      \\
    \end{pmatrix}\,,
\end{align}
which translates within our \(\lambda_k\) basis into the component vector
\begin{align}
    \rho_0^e &= \frac{1}{\sqrt{2N}} \\
    \rho_1^e &= \frac{1}{2}\sin(2\theta) \\
    \rho_2^e &= \ldots = \rho_{N(N-1)}^e = 0 \\
    \rho_{N(N-1) + 1}^e &= \frac{1}{2}\cos(2\theta) \\
    \rho_{N(N-1) + k}^e &= \frac{1}{\sqrt{2k(k+1)}}\,, \quad 2 \leq k \leq N-1\,.
\end{align}
For the other initial neutrino state (the muon and tau superposition \(\nu_{\mu\tau}\)) only the sign of \(\rho_2\) and \(\rho_{N(N-1)+1}\) has to be flipped.
In order to calculate, for example, \(P_{ee}(L)\) we have to find an analytical expression of the trace
\begin{align}
    P_{ee}(L) &= \mathrm{Tr}(\varrho^e\varrho^e(L)) \\
    &= 2 \langle \varrho^e, \varrho^e(L) \rangle \\
    &= 2 \vec{\varrho}^{\,eT} \vec{\varrho}^{\,e}(L) \\
    &= 2 \vec{\varrho}^{\,eT} \exp(\Lambda L) \vec{\varrho}^{\,e}\\
    &= 2 \vec{\varrho}^{\,eT} S \exp(\tilde{D} L) S^{-1} \vec{\varrho}^{\,e}\,.
\end{align}
Using the special block diagonal shape of \(\Lambda\) and the fact that \(\rho_{k}^{e/\mu\tau} = 0\) for \(k = 2, \ldots, N(N-1)\), we can evaluate this expression to
\begin{align}
    P_{ee}(L) =& 2 \left(\frac{1}{2N}+\vec{\rho}_{12}^{\,T} S_{12}^{\nu} \exp\left(\tilde{D}_{12}^{\nu} L\right) S_{12}^{\nu,-1} \vec{\rho}_{12} + \frac{e^{-\Gamma L}}{4}\cos^2(2\theta) \right.\nonumber\\
    &\left.+ \sum_{k = 2}^{N-1} \frac{e^{-\Gamma L}}{2k(k+1)} \right)\,,\\
    \vec{\rho}_{12}^T :=& \rvec{\frac{1}{2}\sin(2\theta)}{0}\,.
\end{align}
After the index \(k = N(N-1)\) the Lindblad operator is purely diagonal resulting in the last two terms of the expression above.
Consequently, we only need to diagonalize the \(\Lambda_{12}^{\nu}\) matrix corresponding to the two neutrino system which is an easy task resulting in
\begin{align}
    \tilde{D}_{12}^{\nu} &= \mathrm{diag}\left(-\frac{1}{L_{12}}-\frac{\Gamma}{2}-i\omega,-\frac{1}{L_{12}}-\frac{\Gamma}{2}+i\omega\right)\\
    \omega &= \sqrt{(\Delta E_{12})^2-\left(\frac{\Gamma}{2}\right)^2}\\
    S_{12}^{\nu} &= \begin{pmatrix}
        -\frac{2i\omega-\Gamma}{2\Delta E_{12}} & \frac{2i\omega+\Gamma}{2\Delta E_{12}}\\
        1 & 1
    \end{pmatrix}\,.
\end{align}
From this, one can straightforwardly deduce the matrix exponential of \(\Lambda_{12}^{\nu}\), i.e.
\begin{align}
    \exp(\Lambda_{12}L) &= e^{-\left(\frac{1}{L_{12}}+\frac{\Gamma}{2}\right)L}\begin{pmatrix}
        \cos(\omega L) + \frac{\Gamma}{2\omega}\sin(\omega L) & -\frac{\Delta E_{12}}{\omega}\sin(\omega L) \\
        \frac{\Delta E_{12}}{\omega}\sin(\omega L) & \cos(\omega L) - \frac{\Gamma}{2\omega}\sin(\omega L)
    \end{pmatrix} \,,
\end{align}
leading to the final result
\begin{align}
    P_{ee}(L) = &2 \left(\frac{1}{2N} + \frac{1}{4}\sin^2(2\theta)e^{-\left(\frac{1}{L_{12}} + \frac{\Gamma}{2}\right) L} \left[\cos(\omega L)+\frac{\Gamma}{2\omega}\sin(\omega L)\right]\right.\nonumber\\
    &+ \left. \frac{1}{4}\cos^2(2\theta)e^{-\Gamma L} + \sum_{k = 2}^{N-1} \frac{1}{2k(k+1)}e^{-\Gamma L}\right)\\
    = &\frac{1}{N} + \frac{1}{2} \sin^2(2\theta)e^{-\left(\frac{1}{L_{12}} + \frac{\Gamma}{2}\right) L} \left[\cos(\omega L)+\frac{\Gamma}{2\omega}\sin(\omega L)\right] \nonumber\\
    &+ \frac{1}{2}\cos^2(2\theta)e^{-\Gamma L} + \frac{N-2}{2N}e^{-\Gamma L}\,.
\end{align}
Here, we used
\begin{align}
    \sum_{k = 2}^{N-1} \frac{1}{k(k+1)} &= \sum_{k = 2}^{N-1} \left(\frac{1}{k}-\frac{1}{k+1}\right) = \frac{1}{2}-\frac{1}{N} = \frac{N-2}{2N}\,.
\end{align}
As usual in the two neutrino case \(P_{ee} = P_{(\mu\tau)(\mu\tau)}\) and \(P_{e(\mu\tau)} = P_{(\mu\tau)e}\) hold.
In order to obtain the oscillation probabilities \(P_{e(\mu\tau)}\) and \(P_{(\mu\tau)e}\), we need to flip the signs in front of the \(\sfrac{1}{2}\sin^2(2\theta)\) and \(\sfrac{1}{2}\cos^2(2\theta)\)
terms.
Lastly, defining \(\gamma := \sfrac{\Gamma}{2}\) concludes our proof of Eqs.~\eqref{eq:mod_result1} and~\eqref{eq:mod_result2}. \(\Box\)

\end{document}